\documentclass[aip,pof,preprint]{revtex4-1}                  
             
\usepackage{graphicx}
\usepackage{amssymb}
\usepackage{amsmath}
\usepackage{subfigure,wrapfig}
\usepackage{ulem}
\usepackage{cancel}
\usepackage[usenames]{color}
\usepackage{hyperref}

\begin{document}

\title{Vortex motion around a circular cylinder above a plane}

\author{G.~L.~Vasconcelos}\email[Corresponding author. Electronic mail: ]{giovani.vasconcelos@ufpe.br}
\affiliation{Laborat\'orio de F\'{\i}sica Te\'orica e Computacional, 
Departamento de F\'{\i}sica, Universidade Federal de Pernambuco,
50670-901, Recife, Brazil.}
\author{M.~Moura }
\affiliation{PoreLab, Department of Physics, University of Oslo, PO Box 1048, Blindern, N-0316, Oslo, Norway.}

\date{\today}

\begin{abstract}
The study of vortex flows around  solid obstacles is of considerable interest from  both a theoretical and practical perspective. One geometry that has attracted renewed attention recently  is that of vortex flows past a circular cylinder placed above a plane wall, where  a stationary recirculating eddy can form  in front of the cylinder, in contradistinction to the usual case (without the plane boundary) for which a vortex pair appears behind the cylinder. Here we analyze the problem of vortex flows past a  cylinder near  a  wall through the lenses of  the point-vortex model. By conformally mapping the fluid domain onto an annular region in an auxiliary complex plane, we  compute the vortex Hamiltonian analytically in terms of certain special functions related to  elliptic theta functions. A detailed analysis of the equilibria of the model is then presented. The location of the equilibrium in front of the cylinder is shown to be in qualitative agreement with  recent experimental findings. We also show that  a topological transition occurs in  phase space as the parameters of the systems are varied.
\end{abstract}


\maketitle

\section{Introduction}
\label{intro}

The formation of vortices in viscous flows past cylindrical structures  is a problem of considerable  interest  from both a theoretical  and applied  perspective, as this process is relevant for  many practical situations, such as vortex-induced vibrations and  stability of submerged structures \cite{sf,vortexvib,mitpaper}. The beauty and intricacies of vortex formation around solid obstacles are well on display in the classical problem of  a flow past a circular cylinder, where  a pair of counter-rotating vortices forms behind the cylinder at small Reynolds numbers, which then goes unstable at higher Reynolds numbers and eventually evolves into a von K\'arm\'an vortex street.  This  system was first studied analytically by F\"oppl in 1913 \cite{foeppl} by considering the motion of  a point vortex around a cylinder  placed in an inviscid and irrotational flow. Since then, the point-vortex model  has  become an important tool to study  vortex dynamics around obstacles, in great part because the model is  amenable to analytical treatment and can be analyzed with standard methods from nonlinear dynamics. In this context, it is worth pointing out the  F\"oppl system has been reanalyzed by several authors  \cite{smith,tangaubry,us2011} and new  dynamical features have recently been discovered  \cite{us2011} in this century-old problem. The question of multiple vortices moving around a cylinder has also been extensively investigated \cite{marsden, shashi2006, borisov2007,us2013} within the point-vortex model.

The  motion of point vortices in the presence of multiple obstacles is also of great interest but here the problem is much more mathematically challenging  because the flow domain is inherently multiply connected. Although many experiments and numerical  studies  of vortex flows around two or more circular cylinders are found in the literature (see, e.g., Ref.~[12] for a brief review of the recent literature),  theoretical analyses  for such cases are much more sparse. Johnson and McDonald \cite{robb2004} studied the motion of a vortex  near two cylinders  (with and without an imposed background flow) using both a point-vortex model and a vortex-patch approach. This study was generalized by Marshall and Crowdy \cite{crowdy2005}  for the case  of a point-vortex in the presence of an arbitrary number of  cylinders (with no  background flow) using conformal mapping techniques based on the Schottky-Klein (SK) prime functions. The formalism of the SK prime functions was also used by Sakajo \cite{sakajo} to study 
the  motion of point vortices in a multiply connected domain  consisting of many circular obstacles inside the unit circle. 

There has also been considerable interest in vortex flows past a circular cylinder near a plane wall; see, e.g., Refs.~[16--20] 
and references therein.   One interesting aspect of this geometry is that  a recirculating eddy can form {\it upstream} of the cylinder as observed by Lin {\it et al.} \cite{lin},  in contrast to  the usual F\"oppl system where the vortices appear behind the cylinder. This observation of a stationary vortex upstream of the cylinder has formed part of the motivation for the present work. 

Here we study this problem in more detail by considering   the motion of a point vortex  in a uniform stream past a cylinder above a plane.  By mapping the physical flow domain  onto an annulus in an auxiliary complex plane, we obtain the Hamiltonian of the model and  study in detail its equilibrium points, with particular emphasis on the fixed point in front of the cylinder. It is shown that this equilibrium is neutrally stable and hence accessible to experimental realization, as was indeed verified by  Lin {\it et al.} \cite{lin}. Furthermore, the locus of these equilibria in the model is in qualitative agreement with the location of the center of the recirculating eddy observed in the  experiments  \cite{lin}. Other aspects of the experiments, such as the fact that  the center of the recirculating eddy moves closer to the cylinder as the  gap between the cylinder and the plane decreases,  can also be explained by the point-vortex model, as will be shown later. Another interesting dynamical feature of our point-vortex system  is that it exhibits  a topological transition in the phase space as  the gap or the vortex strength is varied,   implying  that the nature of the  vortex trajectories can vary drastically as the parameters of the problem crosses certain critical values.

The paper is organized as follows. In Sec.~\ref{sec:CP}, the problem of a point vortex in a uniform stream past a circular cylinder above a plane is formulated and both the instantaneous complex potential and the vortex Hamiltonian  are computed analytically.  The equilibrium points of the system and their associated separatrices are studied in detail in Sec.~\ref{sec:EC}. We conclude in Sec.~\ref{sec:DC} with a summary of our main results and a discussion of their physical relevance, and we  briefly comment on possible extensions of the analysis reported here to the case of vortex motion around multiple cylinders.

\section{Problem Formulation}
\label{sec:CP}

\begin{figure}[t]
\includegraphics[width=0.8\linewidth]{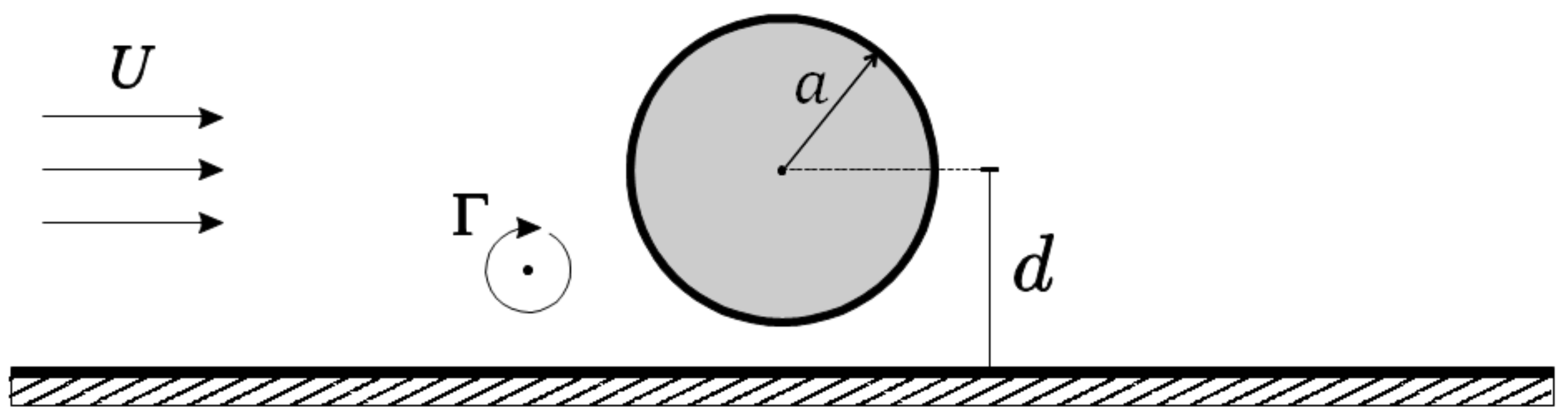}
\caption{Point vortex in a uniform stream past a circular cylinder above a plane.}
\label{fig:cylinder_wall}
\end{figure}

We consider the motion of a point vortex of circulation $\Gamma$ placed in a uniform stream of velocity $U$ past a circular cylinder of radius $a$, whose center lies at a distance $d$ from a plane wall taken to be at $y=0$, so that the gap  between the plane and the cylinder is $\Delta = d-a$; see Fig.~1 for a schematic of the problem geometry.  It is assumed  that  the fluid is inviscid,  irrotational, and incompressible and that the flow has translational symmetry in the direction of the cylinder axis, chosen to be the $z$-axis. Under these assumptions, the fluid velocity field  $\vec{v}(x,y)$ is two-dimensional  and given by the gradient of a potential function:  $\vec{v}=\vec{\nabla}\phi$, where  the velocity potential $\phi(x,y)$ obeys Laplace equation,   $\nabla^2\phi=0$. It is then convenient  to introduce the complex potential $w(z)=\phi(x,y)+i\psi(x,y)$, where  $z=x+\mathrm{i}y$ and $\psi(x,y)$ is the associated stream function.  

\subsection{The complex potential}

\begin{figure}[t]
\includegraphics[width=0.8\linewidth]{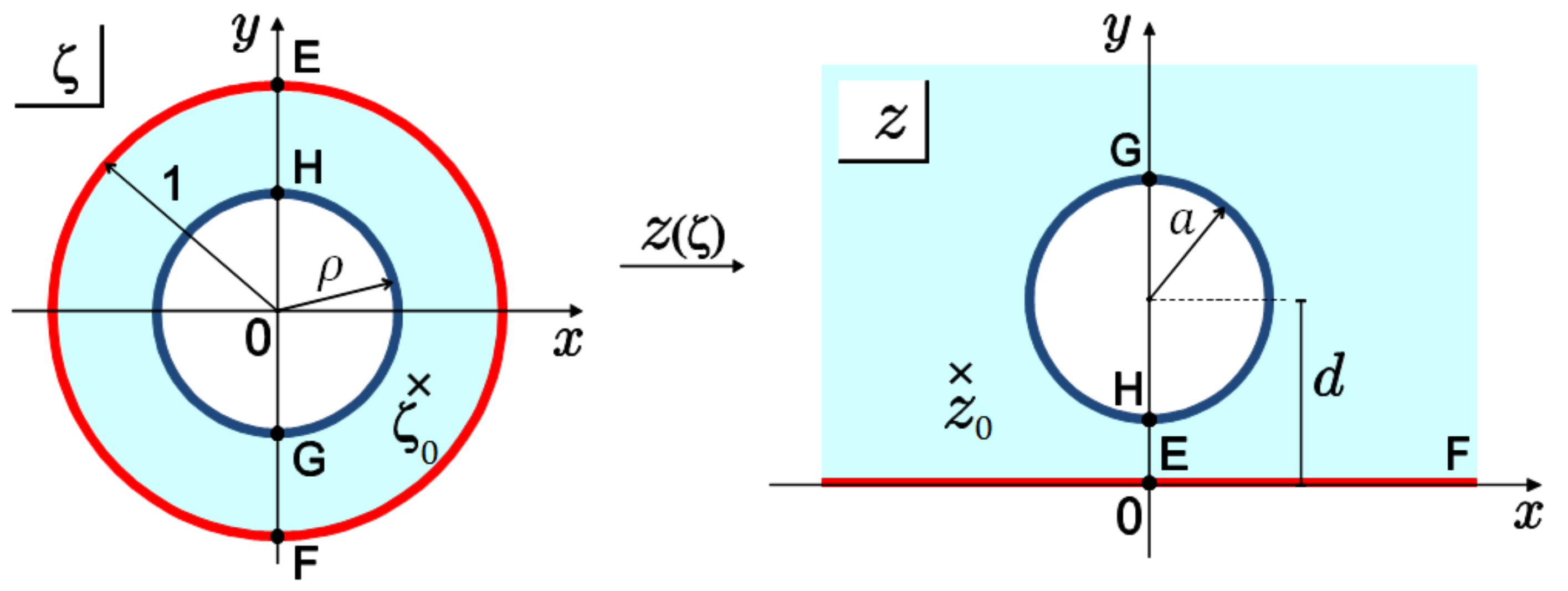}
\caption{Circular domain in the auxiliary complex $\zeta$-plane (left) and the physical domain in the physical $z$-plane (right).}
\label{fig:mapscheme}
\end{figure}

To construct the desired complex potential $w(z)$, we introduce a conformal mapping $z(\zeta)$ from an annular region, $\rho<|\zeta|<1$, in the auxiliary complex $\zeta$-plane onto the fluid domain in the physical $z$-plane; see Fig.~\ref{fig:mapscheme}. The inner circle  ($|\zeta|=\rho$) is chosen to map  to the cylinder, whilst the unit circle ($|\zeta|=1$) is mapped to the plane boundary ($y=0$), with the points $\zeta=i$ and $\zeta=-i$ being mapped to $z=0$ and $z=\infty$, respectively.  The preimage of the  vortex position $z_0$ is denoted by $\zeta_0$. It is not difficult to see that the function that enacts the desired mapping is given by the following M\"obius transformation:
\begin{align}
z(\zeta)=-i\sqrt{d^2-a^2}\,\left(\frac{\zeta-i}{\zeta+i}\right),
\label{eq:z}
\end{align}
whose inverse is
\begin{equation}
\zeta(z)=-i \left(\frac{z-i\sqrt{d^2-a^2}}{z+i\sqrt{d^2-a^2}}\right) .
\label{eq:zeta}
\end{equation}
The radius  $\rho$ of the inner circle in the $\zeta$-plane is related to the physical parameters $a$ and $d$ by the following expression: 
\begin{equation}
\rho=\frac{1-\sqrt{\frac{d-a}{d+a}}}{1+\sqrt{\frac{d-a}{d+a}}} .
\label{r0}
\end{equation}

Next, we introduce the function 
\begin{align*}
W(\zeta)\equiv w(z(\zeta)),
\end{align*}
which corresponds to the complex potential in the auxiliary $\zeta$-plane. The function $W(\zeta)$ can  be written as the sum of two contributions:
\begin{equation}
W(\zeta)= W_U(\zeta)+ \Gamma G_0(\zeta;\zeta_0) ,
\label{eq:W}
\end{equation}
where $W_U(\zeta)$ is the complex potential of a dipole at $\zeta=-\mathrm{i}$, corresponding to a uniform flow of speed $U$ in the  $z$-plane,  and $G_0(\zeta;\zeta_0)$ is the potential for a point vortex of  unit circulation at position $\zeta=\zeta_0$. 

The complex potentials  for both a point vortex and a dipole in an annulus have been obtained by several authors. For example, Johnson and McDonald \cite{robb2004} mapped the annulus to a rectangle and then computed both  potentials in terms of  elliptic  functions. Crowdy and Marshall  \cite{crowdymarshall}  used the properties of the SK prime functions to obtain the complex potential for a point vortex in a multiply connected circular domain of arbitrary connectivity, of which the annulus is  a particular case.  Using the same formalism, Crowdy \cite{crowdy2006} obtained the complex potential for a  uniform  flow past multiple cylinders, which entails computing the complex potential for a dipole in a  multiply connected circular region. 

Following Crowdy and Marshall \cite{crowdymarshall} and Crowdy \cite{crowdy2006},  
one finds that the relevant potentials for our case \cite{us_rcf} are given by 
\begin{align}
G_0(\zeta;\zeta_0)=\frac{1}{2\pi i}\log \left[ \frac{|\zeta_0| P(\zeta/\zeta_0,\rho)}{P(\bar\zeta_0\zeta,\rho)}\right]
\label{eq:G0}
\end{align}
and 
\begin{align}
W_U(\zeta)=-2 \pi U i\sqrt{d^2-a^2} \left.\left(\frac{\partial G_0}{\partial \bar{\zeta_0}}-\frac{\partial G_0}{\partial \zeta_0}\right)\right|_{\zeta_0=-i} ,
\label{eq:WU}
\end{align}
where  $P(x,y)$ is defined by 
\begin{equation}
P(x,y)=(1-x)\prod_{n=1}^{\infty}\left(1-y^{2n}x\right)\left(1-y^{2n}x^{-1}\right).
\label{eq:P}
\end{equation}
The function $P(x,y)$  is related to  the first Jacobi theta function  $\vartheta_1$; see Ref.~[14] for details.

In the problem discussed so far there is no net circulation around the cylinder. A flow with a nonzero circulation $\gamma$ around the cylinder  in the $z$-plane can  be generated by simply adding  to $W(\zeta)$ a term
 of the form  
 \begin{equation}
\frac{\gamma}{2\pi i}\log \zeta,
\label{eq:Wcirc}
\end{equation}
which corresponds in the $\zeta$-plane to a vortex of circulation $\gamma$ at the origin.

Before leaving this section let us establish some auxiliary results that will be needed later. 
First, we  recall   that the so-called {\it  hydrodynamic Green's
function},  here denoted by $G(\zeta,\overline\zeta,\zeta_0,\overline \zeta_0)$, is defined as  the stream function associated with $G_0(\zeta;\zeta_0)$. In other words,
 \begin{align} 
G(\zeta,\overline\zeta,\zeta_0,\overline \zeta_0)={\rm Im}\left[G_0(\zeta,\zeta_0)\right],
\end{align}
which in view of (\ref{eq:G0}) becomes
\begin{align}
{ G}(\zeta,\overline\zeta,\zeta_0,\overline \zeta_0)
&=-\frac{1}{4\pi}\log \left[\frac{\zeta_0\overline \zeta_0P(\zeta/\zeta_0,\rho)P(\overline \zeta/\overline \zeta_0,\rho)}{P(\zeta\overline \zeta_0,\rho)P(\overline \zeta\zeta_0,\rho)}\right],
\label{eq:G}
\end{align}
where  we used  that $\overline{P(\zeta,\rho)}=P(\overline\zeta,\rho)$, with the bar denoting complex conjugation.

Another quantity  that will be needed later is the so-called {\it Robin function}, $\mathcal{R}(\zeta_0,\bar\zeta_0)$, which is given 
by the regular part of $G(\zeta,\overline\zeta,\zeta_0,\overline \zeta_0)$  evaluated at the vortex position:
\begin{align}
\mathcal{R}(\zeta_0,\bar\zeta_0)&=\left[-G(\zeta,\overline\zeta,\zeta_0,\overline \zeta_0)-\frac{1}{2\pi}\log|\zeta-\zeta_0|\right]_{\zeta=\zeta_0}
\label{eq:R}
\end{align}
which in view of  (\ref{eq:G}) becomes
\begin{align}
\mathcal{R}(\zeta_0,\bar\zeta_0)&=\frac{1}{2\pi}\log \left(\frac{P'(1,\rho)}{P(\zeta_0\overline \zeta_0,\rho)}\right),
\end{align}
where $P'(x,y)=P(x,y)/(1-x)$. After disregarding the (irrelevant) constant  term depending  only on $\rho$, the expression above simplifies to
\begin{align}
\mathcal{R}(\zeta_0,\bar\zeta_0)&=-\frac{1}{2\pi}\log \left[{P(\zeta_0\overline \zeta_0,\rho)}\right].
\label{eq:R0}
\end{align}

\subsection{The Hamiltonian}
\label{sec:H}

For convenience of notation,  we shall henceforth drop the subscript from the symbols  denoting the location of the vortex, and so we will use  $z(t)=x(t)+\mathrm{i}y(t)$ to represent the  position of the vortex  in the $z$-plane at  time $t$, while  denoting  the position of the vortex in  $\zeta$-plane simply by $\zeta$.   

As shown by Lin \cite{lin1941},  the motion of point vortices in the presence of  rigid boundaries  can  be  formulated as a Hamiltonian dynamics, with the same canonical structure as in the case without  boundaries.  Let us  first consider the vortex dynamics in the $\zeta$-plane and denote the corresponding Hamiltonian   by $H^{(\zeta)}(\zeta,\overline \zeta)$.
It has  been shown by Lin \cite{lin1941} that  $H^{(\zeta)}(\zeta,\overline \zeta)$  is of the form
\begin{align}
H^{(\zeta)}(\zeta,\overline \zeta)=
\Gamma\psi(\zeta,\overline \zeta)-\frac{1}{2}\Gamma^2{\cal R}(\zeta,\overline\zeta),
\label{eq:HM1}
\end{align}
where $\psi(\zeta,\overline \zeta)$ is the stream function associated with the background flow  and ${\cal R}(\zeta,\overline\zeta)$ is
the  Robin function, which in our case is given by Eq.~(\ref{eq:R0}). The background flow in the $\zeta$-plane is described by the potentials (\ref{eq:WU}) and (\ref{eq:Wcirc}), and so we have 
\begin{align}
 \psi(\zeta,\overline \zeta)
 &={\rm Im}\left[W_U(\zeta) \right] -\frac{\gamma}{4\pi}\log(\zeta\overline\zeta).
 \label{eq:psi0}
\end{align}
Substituting  (\ref{eq:psi0}) and (\ref{eq:R0}) into (\ref{eq:HM1})  yields
\begin{align}
H^{(\zeta)}(\zeta,\overline \zeta)=
\Gamma {\rm Im}\left[W_U(\zeta)\right] -\frac{\gamma\Gamma}{4\pi}\log(\zeta\overline\zeta)-\frac{\Gamma^2}{4\pi}\log \left[{P(\zeta\overline \zeta,\rho)}\right].
\label{eq:HM2}
\end{align}

It   also follows from a general result  due to Lin \cite{lin1941} that the Hamiltonian   $H^{(z)}(z,\overline z)$ in the $z$-plane is given by the following relation
\begin{align}
H^{(z)}(z,\overline z)=H^{(\zeta)}(\zeta,\overline \zeta)+\frac{\Gamma^2}{4\pi}\log\left|z_\zeta(\zeta)\right|,
\label{eq:Hz0}
\end{align}
where   $z_\zeta$ denotes the derivative of the mapping $z(\zeta)$. Now, from (\ref{eq:z}) one finds 
\begin{align*}
z_\zeta(\zeta)=-\frac{2\sqrt{d^2-a^2}}{(\zeta+i)^2},
\end{align*}
 which implies that
\begin{align*}
|z_\zeta(\zeta)|=\frac{2\sqrt{d^2-a^2}}{(\zeta+i)(\overline\zeta-i)},
\end{align*}
and hence (\ref{eq:Hz0}) becomes
\begin{align}
H^{(z)}(z,\overline z)=H^{(\zeta)}(\zeta,\overline \zeta)-\frac{\Gamma^2}{4\pi}\log\left[(\zeta+\mathrm{i})(\overline \zeta-\mathrm{i})\right],
\label{eq:Hz}
\end{align}
where an irrelevant additive constant was omitted.  Inserting (\ref{eq:HM2}) into (\ref{eq:Hz}) and using the inverse mapping (\ref{eq:zeta}),  then gives the desired Hamiltonian $H^{(z)}$ in the $z$-plane.

\section{Equilibrium Configurations}
\label{sec:EC}

Here we wish to  compute the equilibrium points of the system and study how they depend on the parameters of the problem. Without loss of generality,  we shall set  $U=1$ and $a=1$,   so  that we are  left with only three free physical parameters, namely the vortex circulation  $\Gamma$, the gap $\Delta$ between the cylinder and the plane, and  the net circulation $\gamma$ around the cylinder. 

\begin{figure}[t]
\includegraphics[width=0.9\linewidth]{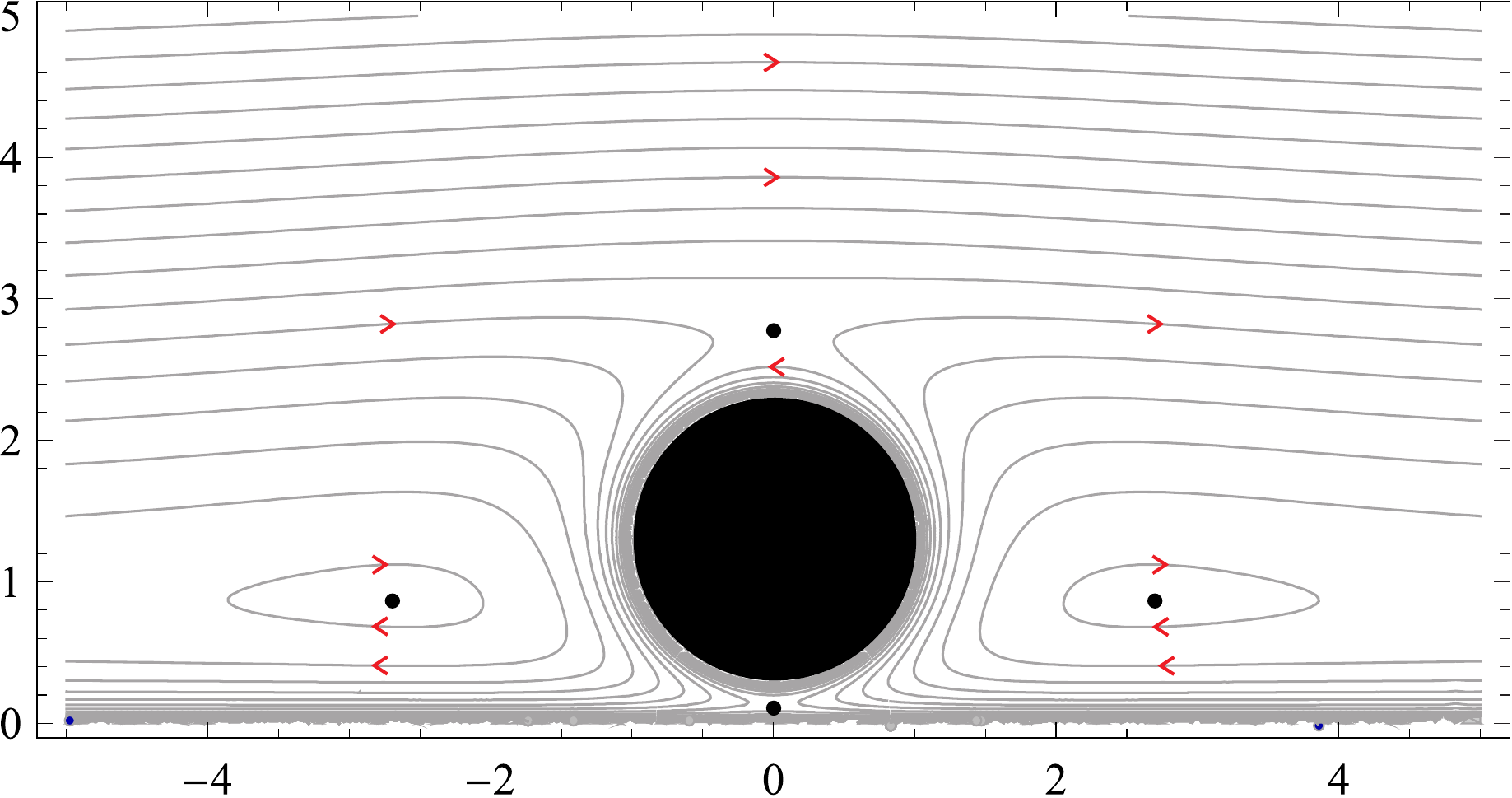}
\caption{Vortex trajectories obtained from a contour plot of the Hamiltonian. The black dots denote the fixed points and the arrows indicate the direction of motion. Here the parameters  are  $\Gamma=-10$, $\Delta=0.3$, and $\gamma=0$. }
\label{fig:contour}
\end{figure}

\subsection{Phase portrait}

As the Hamiltonian is a conserved quantity, the vortex trajectories correspond to level curves of the Hamiltonian $H^{(z)}$. Figure~\ref{fig:contour} shows the phase portrait of the system obtained from  a contour plot of the Hamiltonian given in (\ref{eq:Hz}) for  $\Gamma=-10$, $\Delta=0.3$, and $\gamma=0$.   
From this figure one can immediately identify four equilibrium points  (black dots in the figure): two {\it saddle points} located above and below the cylinder on the axis bisecting the cylinder  perpendicularly to  the oncoming flow; and two {\it centers}, being one behind and the other in  front  of the cylinder. (In Fig.~\ref{fig:contour}  and  subsequent figures,  the function $P(x,y)$  has been calculated by considering terms up to the fifth level, i.e., by computing the  product appearing in (\ref{eq:P}) up to terms with $n=5$. We found that increasing the number of terms does not make any noticeable difference within the scale of our plots.) 

In addition to these four equilibria, there is another fixed point at infinity, namely at 
\begin{equation}
x=\pm\infty, \qquad y_\infty=\frac{\Gamma}{4\pi U} ,
\label{eq:yc}
\end{equation}
which corresponds to the equilibrium for a vortex pair placed in a uniform stream in the absence of boundaries. This equilibrium is analogous to the fixed point  at infinity  that exists in the F\"oppl system \cite{us2011}. As discussed in detail in Ref.~[7],  this equilibrium corresponds to a {\it nilpotent saddle} in the sense that the Jacobian matrix of the linearized system has two zero eigenvalues with identical eigenvectors.

Other general dynamical features  of  our system seen in Fig.~\ref{fig:contour}  are worth commenting on here. First, note that there  is  a set of closed (periodic) orbits encircling each of the equilibria behind and in front of  the cylinder, thus confirming  their elliptic nature. Far above the cylinder, there are open orbits going  from negative infinity to plus infinity,  meaning that in these cases the vortex follows an almost straight trajectory from left to right as it is simply carried away by the uniform stream without being  much affected by the cylinder. On the other hand, a vortex placed very close to the cylinder (or to the plane) will circle around the cylinder (or move parallel to the plane), because it feels very strongly the velocity induced by the nearest images on these boundaries.  The nature of the other  `intermediate'  trajectories  depends on  the strength of the vortex and the gap between the cylinder and the plane, as  will be discussed in Sec.~\ref{sec:tpt}. But before that, let us study in more detail  the equilibria in front of the cylinder.

\subsection{Equilibrium  in front of the cylinder}
\label{sec:efc}

\begin{figure}[t]
\includegraphics[width=0.9\linewidth]{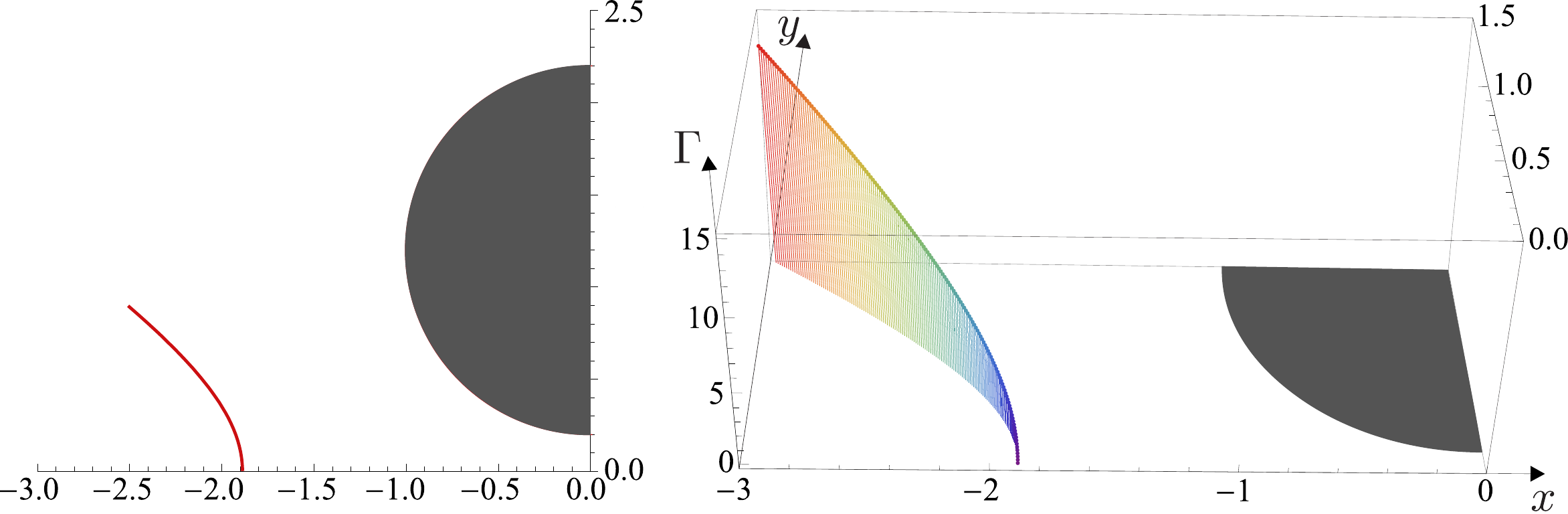}
\caption{Locus of equilibria for $\Delta=0.2$ and $\gamma=0$ (left panel), with the right panel showing in  perspective  the value of the vortex circulation $\Gamma$ (vertical axis) on the curve of equilibria  shown in the left panel.}
\label{fig:locus}
\end{figure}

The exact location of the equilibria of the system can be obtained by finding the extrema of the Hamiltonian:
\begin{align}
\frac{\partial H^{(z)}}{\partial x}=0,\\
\frac{\partial H^{(z)}}{\partial y}=0.
\end{align}
We have solved these equations numerically for the location of the fixed point in front of the cylinder as a function of the  parameters of the problem.  An example of the locus of these equilibria for $\Delta=0.2$  and $\gamma=0$ is shown in the left panel of Fig.~\ref{fig:locus}, while the right panel of this figure shows the corresponding value of the vortex circulation on the equilibrium curve. The equilibrium starts (for $\Gamma=0$) at the real axis and moves away from the cylinder as $|\Gamma|$ increases. For $|\Gamma|\to\infty$ the numerical results seem to indicate that the curve of equilibria approaches a straight line, but we have not been able to prove this analytically.

\begin{figure}[t]
\subfigure[\label{fig:locus_gap}]{\includegraphics[width=0.46\linewidth]{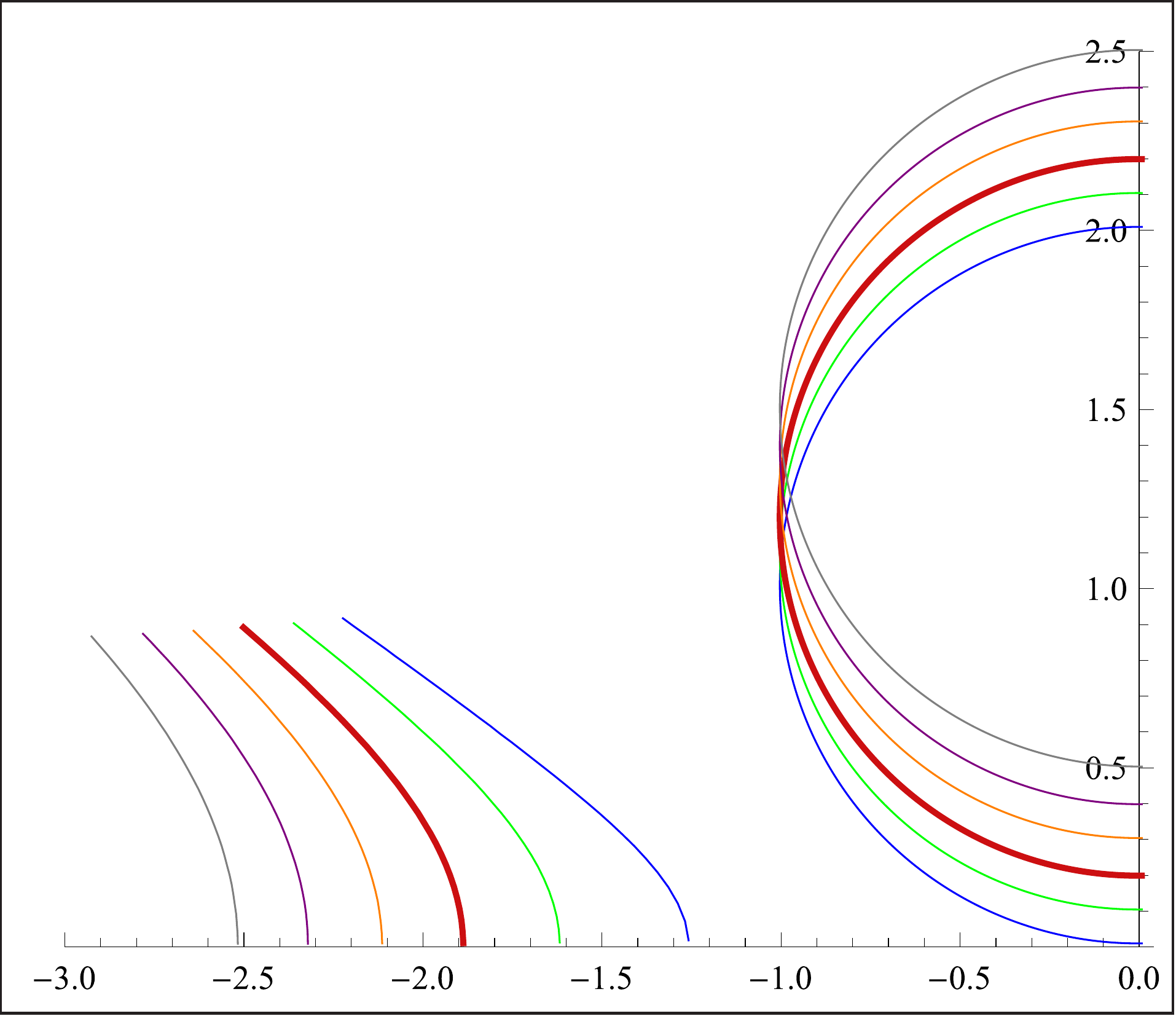}}\qquad
\subfigure[\label{fig:locus_circulation}]{\includegraphics[width=0.4\linewidth]{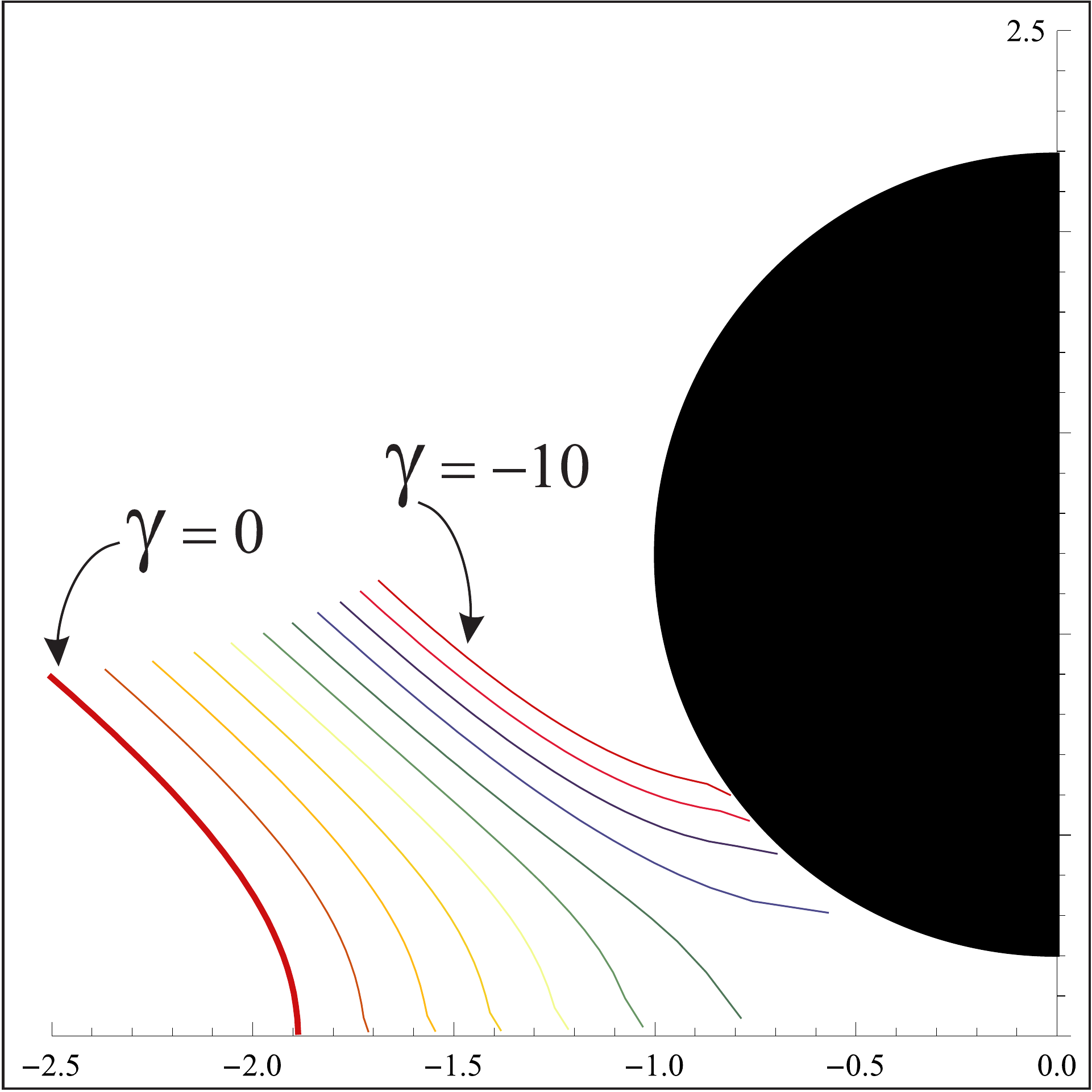}}
\caption{Equilibria in front of the cylinder. (a) Loci of equilibria  for $\gamma=0$ and $\Delta=0.01, 0.1, 0.2, 0.3, 0.4, 0.5$  (from right  to left). (b) Loci of equilibria for $\Delta=0.2$ and $\gamma=0, -1, ...,-10$.  On each curve the vortex circulation varies from $\Gamma=-0.1$ (closer to the plane) to $\Gamma=-10$ (away from the plane).}
\end{figure}

As the gap decreases (increases),  the locus of equilibria moves towards  (away from) 
the cylinder as shown in Fig.~\ref{fig:locus_gap}. This  is in agreement with in the experiments performed by Lin {\it et al.} \cite{lin} where  it was found
that the center of the recirculating eddy moves closer to the cylinder as the gap decreases. However,   the vertical positions of the equilibria shown in  Fig.~\ref{fig:locus_gap} remain approximately the same irrespective of the gap, whereas in the experiments  the  distance from the center of the recirculating eddy to the plane boundary increases more significantly as the gap decreases. This suggests that one needs to include the effect of circulation around the cylinder if one wishes to get  a better qualitative agreement between  the model and the experiments.

In Fig.~\ref{fig:locus_circulation} we show the loci of equilibria for  $\Delta=0.2$ and different values of the circulation $\gamma$ around the cylinder. One sees from this figure that the equilibrium  moves away from  the plane $y=0$  and towards the cylinder as more (negative) circulation is added to the cylinder. As mentioned above, a similar behavior   for the location of the recirculating eddy was found  in the experiments \cite{lin} as the gap was decreased. Our results thus seem to indicate that, insofar as the location of the recirculating eddy is concerned,  the most relevant effect of decreasing the gap in the experiments is a corresponding increase in the clockwise circulation around the cylinder, which  tends to bring the vortex closer to the cylinder.

\begin{figure}[t]
\includegraphics[width=0.5\linewidth]{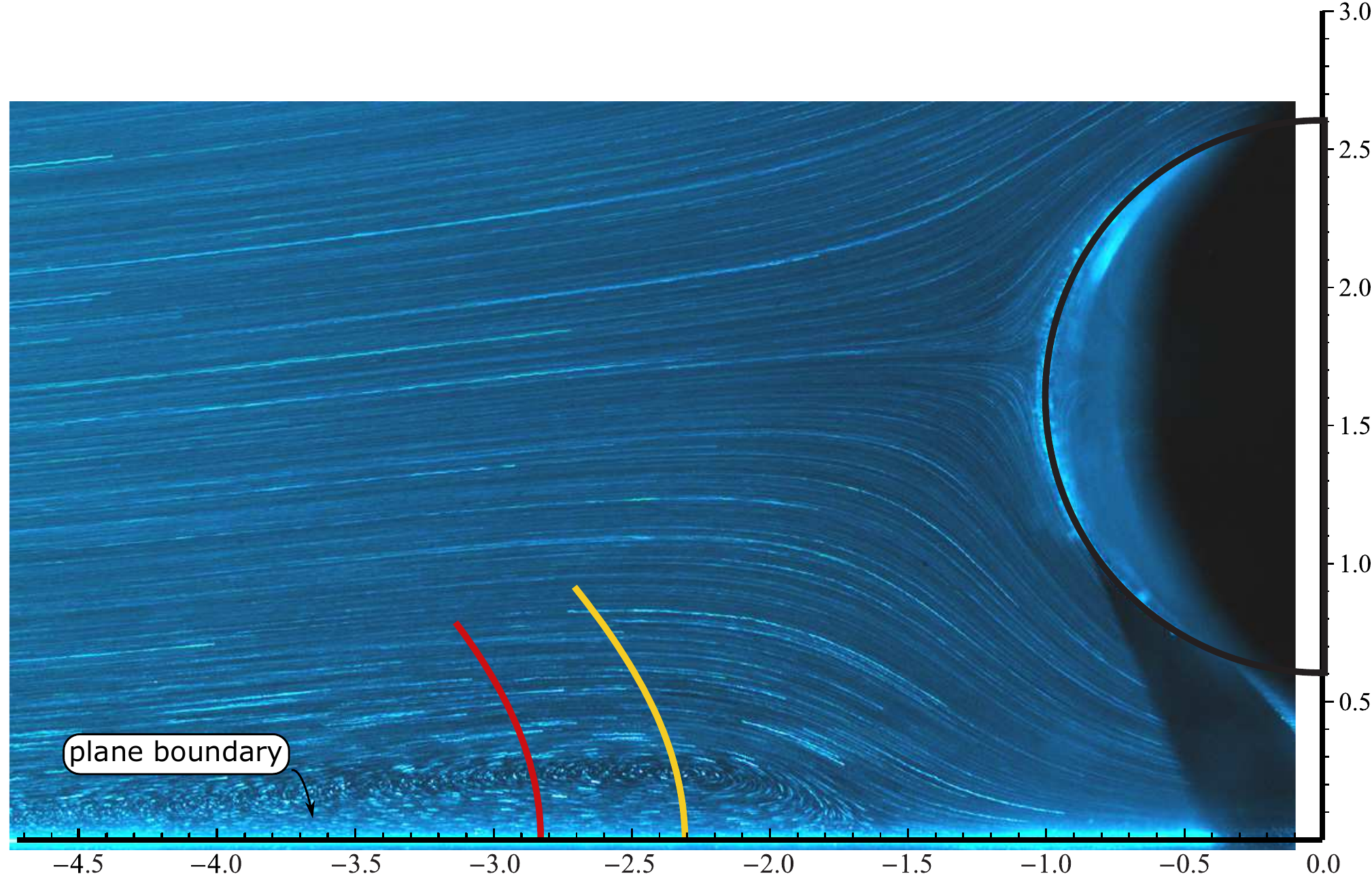}
\includegraphics[width=0.5\linewidth]{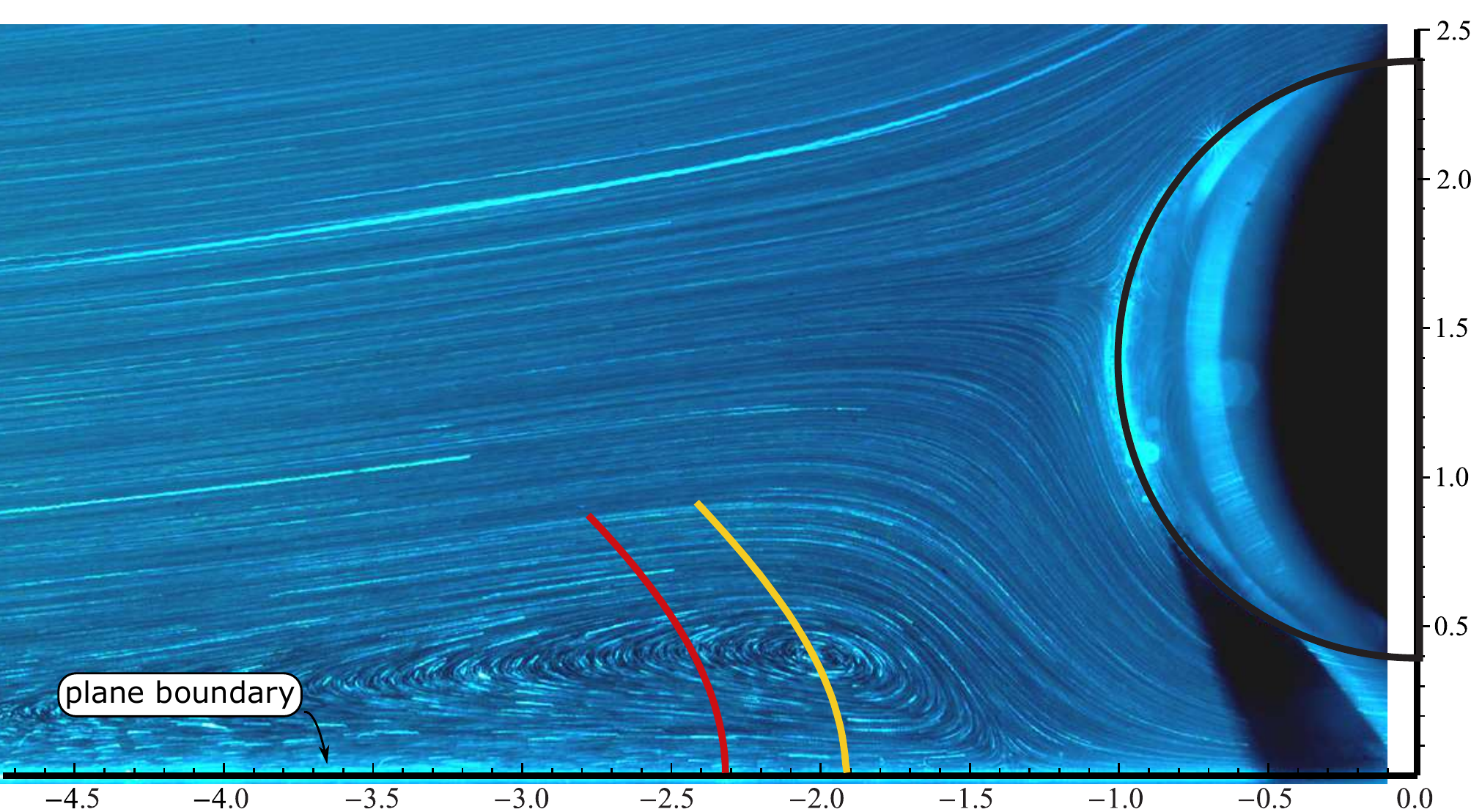}
\includegraphics[width=0.5 \linewidth]{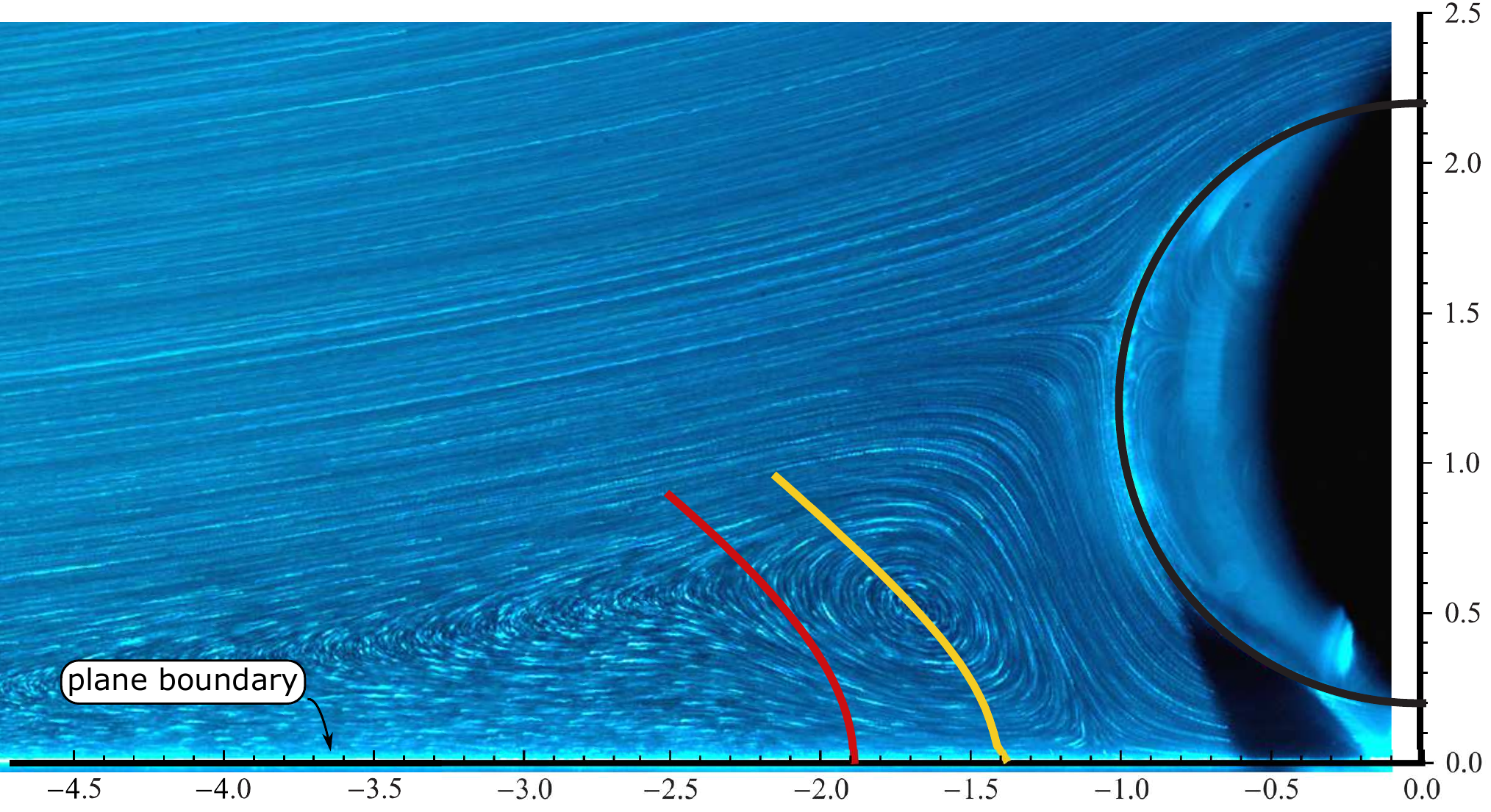}
\caption{Comparison with experiments performed by Lin {\it et al.} \cite{lin} for different gaps:  (a) $\Delta=0.6$,   (b) $\Delta=0.4$,   and (c)  $\Delta=0.2$ . The curves  on the left  (red)  are the loci of equilibria predicted by the model with no circulation around the cylinder ($\gamma=0$), whereas the curves on the right  (yellow)  are for $\gamma=-1.1$ (a),  $\gamma=-1.6$ (b), and $\gamma=-3.0$ (c). The centers $z_0$ of the recirculating eddies and the corresponding  vortex circulations $\Gamma_0$ at these points are as follows:  $z_0 = -2.34 + 0.24 i$ and $\Gamma_0=-2.5$ (a);   $z_0=-2.02 + 0.41 i$ and $\Gamma_0=-4.0$ (b); and   $z_0=-1.69+0.52  {i}$ and $\Gamma_0=-4.5$ (c).  Photographs courtesy of C.~Lin.}
\label{fig:comparison}
\end{figure}

A more direct comparison between the model and the experiments is given in Fig.~\ref{fig:comparison}, where we show pictures  of the flow from the experiments by  Lin {\it et al.} \cite{lin} for three values  of the gap: $\Delta=0.6$  (a), 0.4 (b), 0.2 (c).  The red curves on the left of each  figure indicate the  loci of equilibria predicted by the model  for the case of zero circulation ($\gamma=0$) around the cylinder.  As one can see from the figures,  the centers of the recirculating eddies do not fall on the theoretical curves of equilibria with $\gamma=0$, and hence nonzero circulation around the cylinder needs to be taken into account.  
For each case shown in Fig.~\ref{fig:comparison}, we  then determined the  circulation $\gamma$ around the cylinder for which the theoretical locus of equilibria (yellow curves on the right)  passes   through the center $z_0$ of the recirculating eddy, and we also recorded the vortex circulation $\Gamma_0$ at that point (i.e., where the theoretical equilibrium coincides with the eddy center). The calculated values for each case are as follows: $\gamma=-1.1$, $\Gamma_0=-2.5$, and $z_0=-2.34 + 0.24   {i}$ for $\Delta = 0.6$;  $\gamma=-1.6$, $\Gamma_0=-4.0$, and $z_0=-2.02 + 0.41   {i}$ for $\Delta = 0.4$; and  $\gamma=-3.0$, $\Gamma_0=-4.5$, and $z_0=-1.69+0.52  {i}$ for $\Delta = 0.2$.

Several comments  about these results are in order. First,  note that the point-vortex strength $(|\Gamma_0|$) for the equilibrium at the center of the recirculation zone increases as the gap decreases. This is in agreement with the experimental finding  that ``the size of the recirculating eddy increases with the decrease in gap ratio'' \cite{lin}. In other words, the vortex strength in the model is directly related to the size of the observed recirculation zone. The point-vortex model also explains the experimental observation that  the vertical position of the recirculating eddy  increases as the gap decreases---such an upward displacement of the eddy centre  is mainly due to the accompanying increase in circulation around the cylinder, as mentioned above.  In fact, the authors of Ref.~[18] implicitly noticed this connection when they argued that reducing the gap causes an enlargement of the eddy which ``deflects part of  the fluid  from upstream of the plane boundary over the top of the circular cylinder," thus effectively increasing the circulation around the cylinder. Our analysis above based on the point-vortex model shows that this is indeed the case: the circulation around the cylinder increases from 1.1 to 3.0 (in magnitude) as the gap decreases from 0.6 to 0.2.
It should be pointed out, however, that since additional experimental parameters of the flow  are not known, such as the effective vortex strength and the circulation around the cylinder,
there is admittedly some degree of  flexibility in our fitting of the center of the  recirculation zone with the point-vortex  model.  But notwithstanding this caution, the model  does provide a coherent physical framework in which the experimental observations can be understood, as shown above. Further comparison between theory and experiment is discussed below.

\begin{figure}[t]
\vspace{-3cm}
\includegraphics[width=0.5\linewidth]{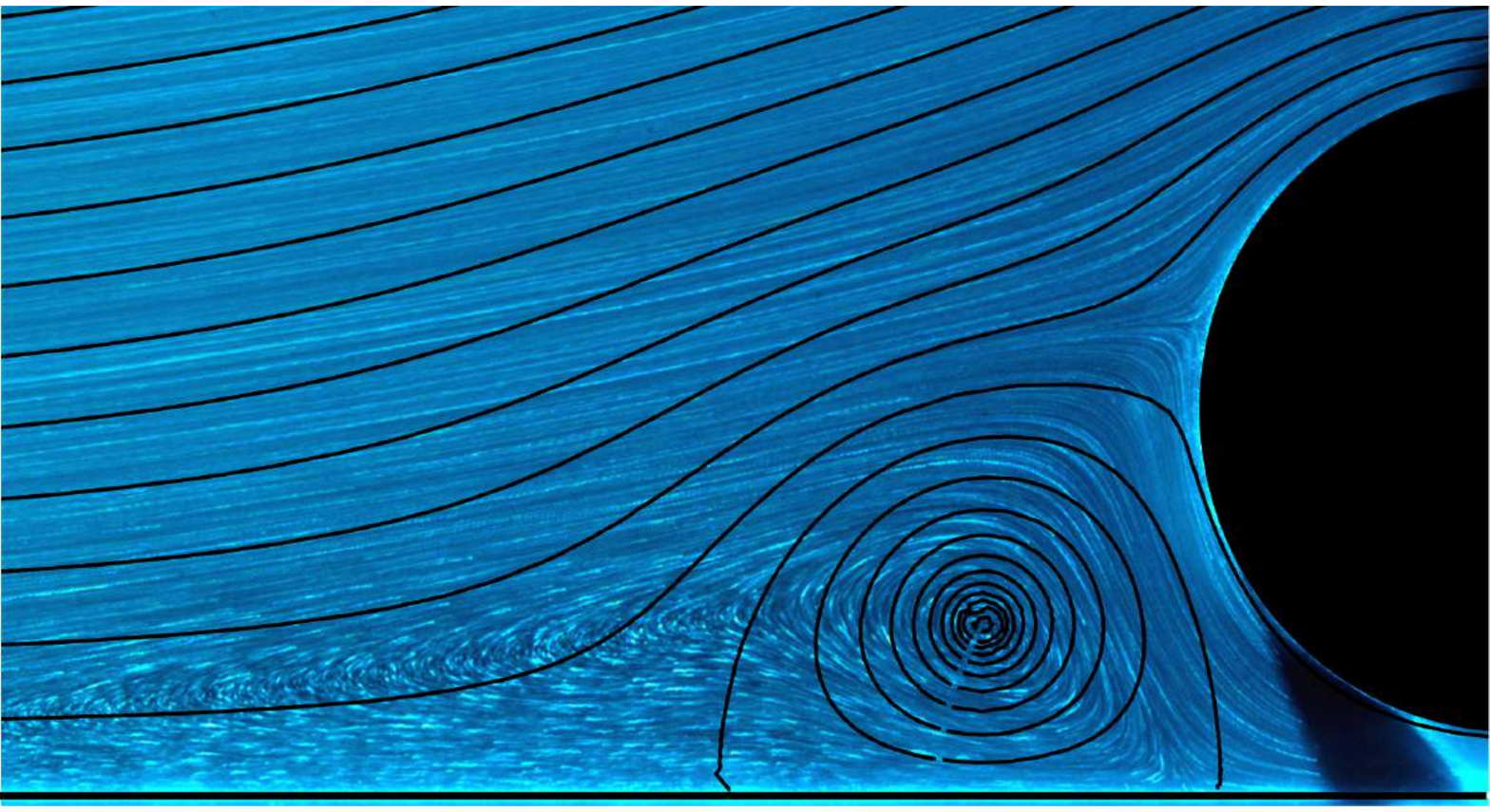}
\vspace{-3cm}
\caption{Streamlines obtained from the point vortex model superimposed on the  photograph of the  experimental flow shown in Fig.~\ref{fig:comparison}(c). The  parameters are $\Gamma=-4.5$, $\Delta=0.2$, and $\gamma=-3$, with the vortex  at the equilibrium position $z_0=-1.69+0.52\mathrm{i}$. The  faint white streak at the bottom of the streamline plot is an artifact of the plotting routine.}
\label{fig:streamlines}
\end{figure}

In Fig.~\ref{fig:streamlines} we show a set of streamlines obtained from the point-vortex model,
superimposed on the photograph of the flow corresponding to  Fig.~\ref{fig:comparison}(c),
where it is seen that the theoretical streamlines bear a general  resemblance to the experimental flow pattern. Note, in particular, that the model  correctly predicts the  approximate location of the stagnation point in front of the cylinder.  There is also, of course,   noticeable disagreement between theory and experiments, particularly  with respect to the shape of the recirculation zone.
This is not surprising given that  the  point-vortex model is not designed to reproduce the streamline patterns of the real flows it intends to describe. For instance, recirculating eddies in real flows are  quite sensitive to viscous effects  which are not accounted for in the  idealized  model of point vortices in an inviscid fluid. This explains in part the fact that the experimental eddy in Fig.~\ref{fig:streamlines} is considerably elongated in the upstream direction in comparison to the model prediction, which is a consequence of viscous effects near the plane boundary.  Similar effect is seen in the case of a flow past a surface-mounted blunt obstacle, where the  recirculating eddy upstream of the obstacle becomes more elongated with increasing Reynolds number  as a result of the increased viscous effects near the upstream surface \cite{blunt}. 
The discrepancy between theory and experiments regarding the shape of the recirculating eddy  is  also reminiscent of similar behavior observed in an unbounded flow past a cylinder, where the general properties of the vortex pair behind the cylinder are well described by the  F\"oppl point-vortex system \cite{foeppl,us2011}, although the theoretical and experimental streamline patterns differ somewhat  \cite{marcel_thesis}.  In the case of a flow past a cylinder above a plane,  viscous effects are more pronounced owing to the presence of the solid plane wall. Nonetheless,  it is remarkable that the point-vortex  model is capable of explaining most of the experimental findings concerning the equilibria in front of the cylinder, as shown above.

Another  shortcoming of the point-vortex model  is  the fact that it has a  singular vorticity distribution concentrated at isolated points, whereas in real flows the vorticity is distributed over a finite volume. One possible way to desingularize the model is to consider an approach where patches of constant vorticity are embedded in an inviscid and irrotational fluid  \cite{robb2004, elcrat1,elcrat2}. The added complication in this case is that the boundaries of the vortex patches have to be computed as part of the solution. As analytical solutions for such free-boundary problems are rare, particularly if solid surfaces are  present, a numerical treatment is almost always required   to tackle the problem \cite{robb2004}. The point-vortex model, in contradistinction, has the advantage that it is more amenable to analytical treatment while still representing a valid approximation for vortex flows around solid obstacles. In this context, it is  noteworthy to point out that the point-vortex model has also been successfully used to investigate vortex flows in the presence of free surfaces  \cite{fs1,fs2,fs3}. This is another instance where the model captures the physics of the flow while allowing for a simpler mathematical treatment of the problem.

\subsection{Topological transitions}
\label{sec:tpt}

As  discussed above, our system possesses  five equilibrium points: two centers,  two saddles, 
 and a nilpotent saddle point at infinity; see Fig.~\ref{fig:contour}.  The stable and unstable manifolds associated with the saddle points correspond to separatrices of the vortex motion, which play an important role in defining the nature of the orbits in phase space. We recall, in particular, that separatrices  connecting a saddle point either to itself or to another saddle point form a homoclinic or a heteroclinic loop, respectively, so that for any initial condition inside one such loop the vortex remains `trapped' within the loop.

\begin{figure}[t]
{\includegraphics[width=0.7\linewidth]{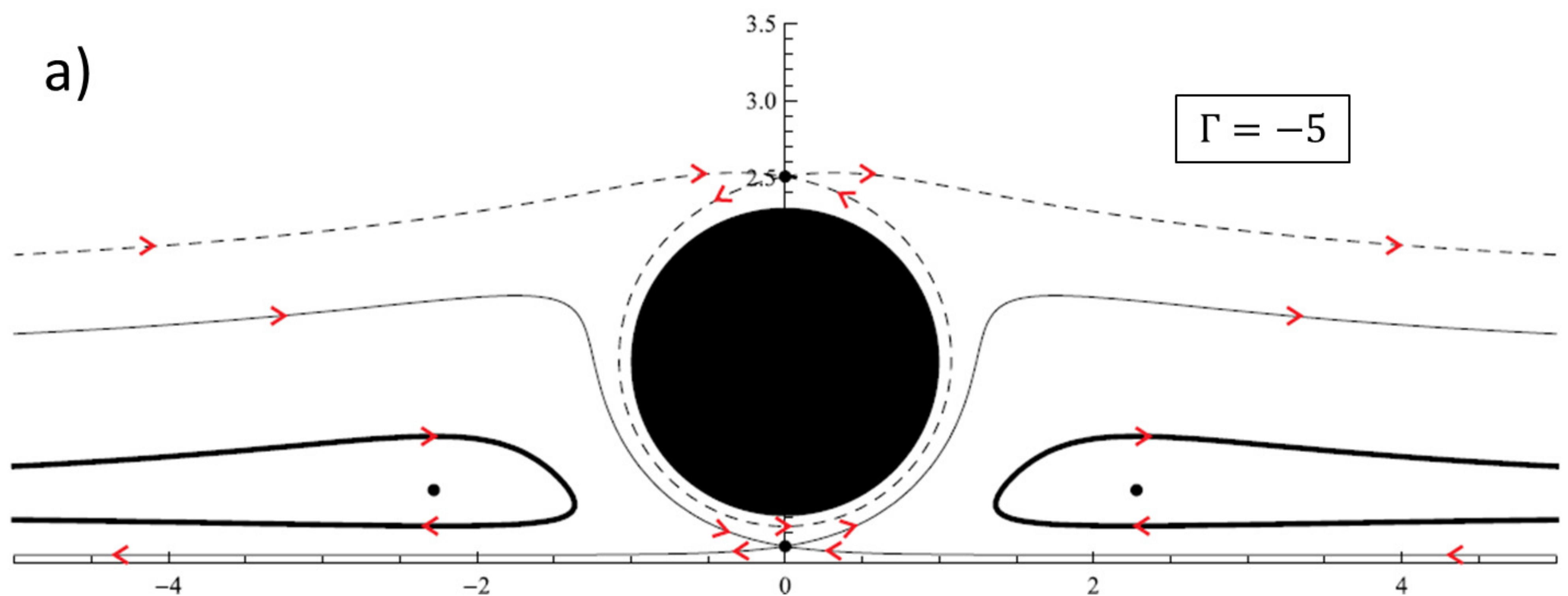}}
{\includegraphics[width=0.7\linewidth]{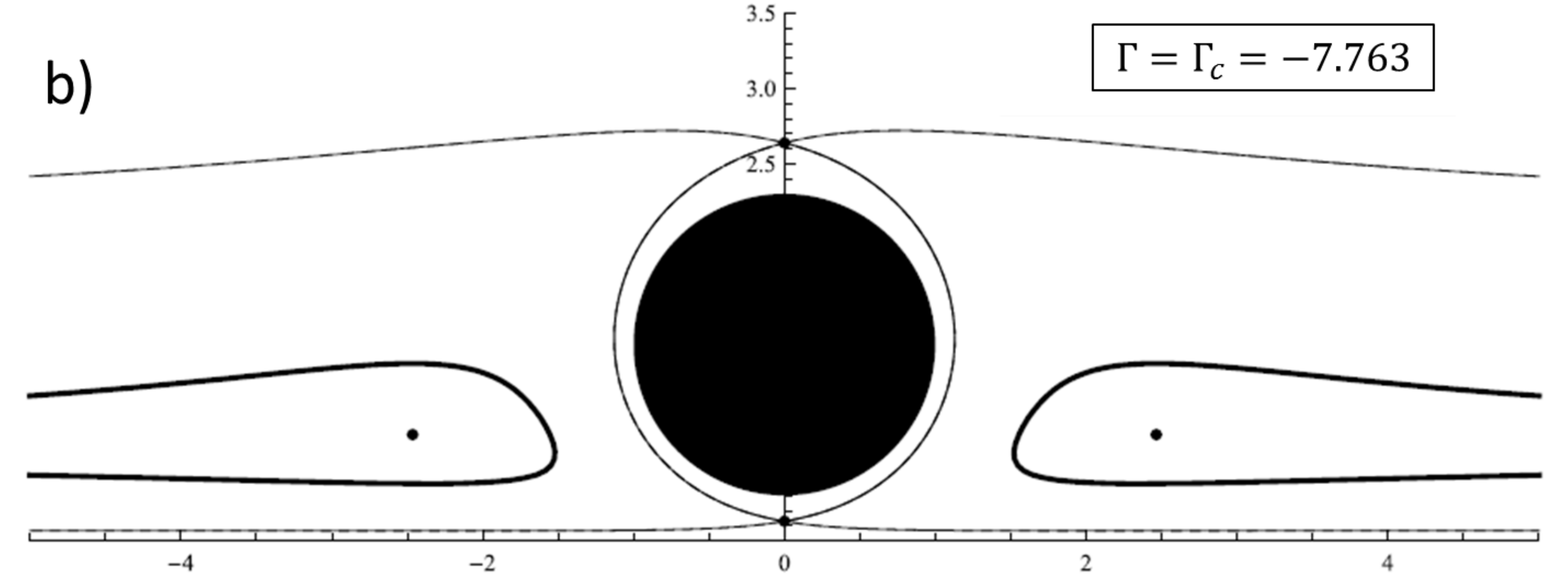}}
{\includegraphics[width=0.7\linewidth]{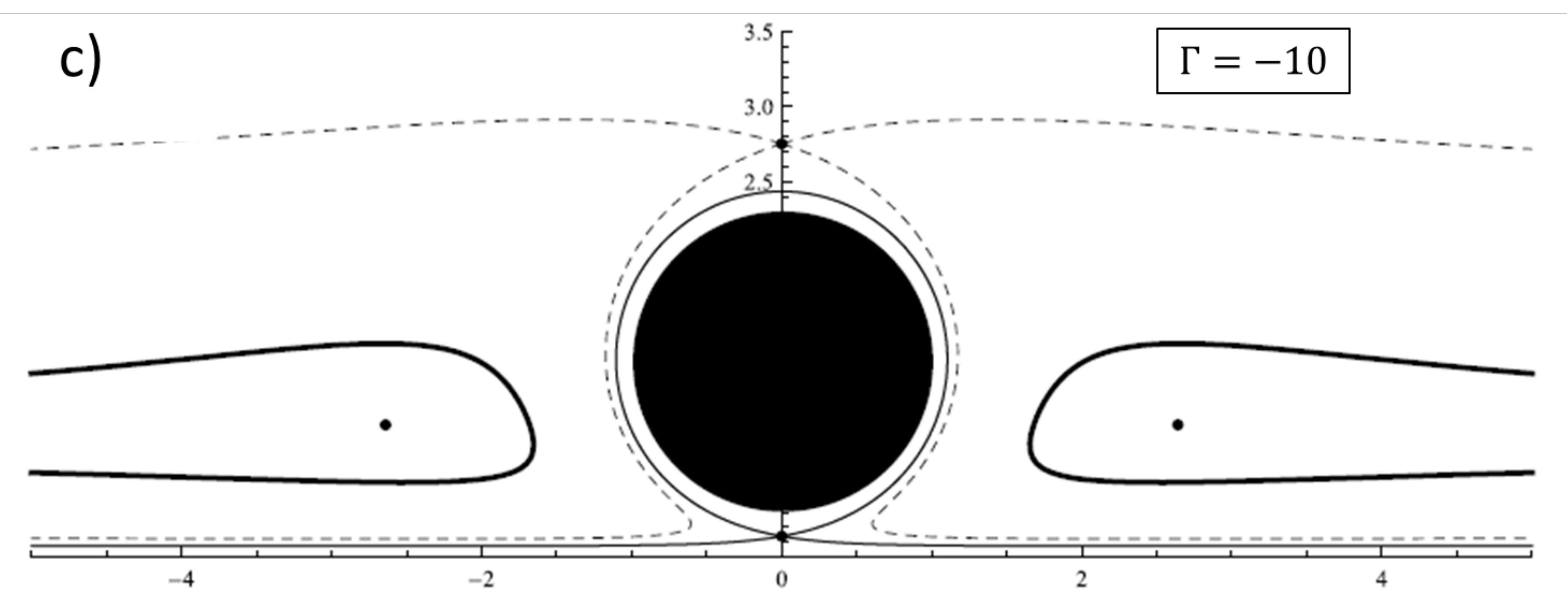}}
\caption{Topological transition in phase space as the vortex circulation is  varied while the gap is kept fixed at $\Delta=0.3$ and $\gamma=0$. The trajectories shown in the figure correspond to the separatrices for: (a)  $\Gamma=-5$, (b) $\Gamma=\Gamma_c=-7.763$ and (c) $\Gamma=-10$.}
\label{fig:sepa1}
\end{figure}

To illustrate the possible topological patterns in phase space,  we show in Fig.~\ref{fig:sepa1}  the  separatrices  for $\Delta=0.3$, $\gamma=0$,  and three different values of $\Gamma$. As seen in this figure, the separatrices (thick solid lines) associated with the nilpotent saddle point at infinity   form two homoclinic loops,  called nilpotent saddle loops  \cite{us2011}, which  encircle the  centers.   These  nilpotent saddle loops define the region of nonlinear stability of the two centers, in the sense that  for any initial condition inside one such  loop the ensuing orbit is closed (i.e., periodic). The separatrices  emanating from the saddle points on the vertical axis can   form either homoclinic or heteroclinic orbits, depending on the value of $\Gamma$.  This is illustrated in Fig.~\ref{fig:sepa1},  where one sees  that for a  critical value of $\Gamma=\Gamma_c=-7.763$ there exists a heteroclinic loop  connecting the two saddles and encircling the cylinder, see Fig.~\ref{fig:sepa1}(b), whereas for $\Gamma\ne \Gamma_c$ the heteroclinic loop breaks up and one  homoclinic loop is formed connecting either the saddle above the cylinder (if $|\Gamma|<|\Gamma_c|$) or the saddle point below  (for $|\Gamma|>|\Gamma_c|$); see Figs.~\ref{fig:sepa1}(a) and \ref{fig:sepa1}(c). This formation and disappearance of loops as the parameter $\Gamma$ is varied   corresponds to  a `topological transition'  in phase space, which  induces a drastic change in the  nature of the vortex trajectories  as the critical value $\Gamma_c$ is crossed. For example, for $|\Gamma|<|\Gamma_c|$ there is a set of trajectories for which the vortex---starting from upstream infinity---will  approach the cylinder, go around it counterclockwise,  and pass between the cylinder and the plane, before moving  off to  infinity downstream; see Fig.~\ref{fig:sepa1}(a). For $|\Gamma|\ge|\Gamma_c|$ no such trajectories exist, meaning that a vortex coming from upstream infinity is  either `reflected' back to  infinity upstream,  or  moves to downstream infinity without being very much affected by the cylinder; see Figs.~\ref{fig:sepa1}(b) and \ref{fig:sepa1}(c). 
 
\begin{figure}[t]
\includegraphics[width=0.7\linewidth]{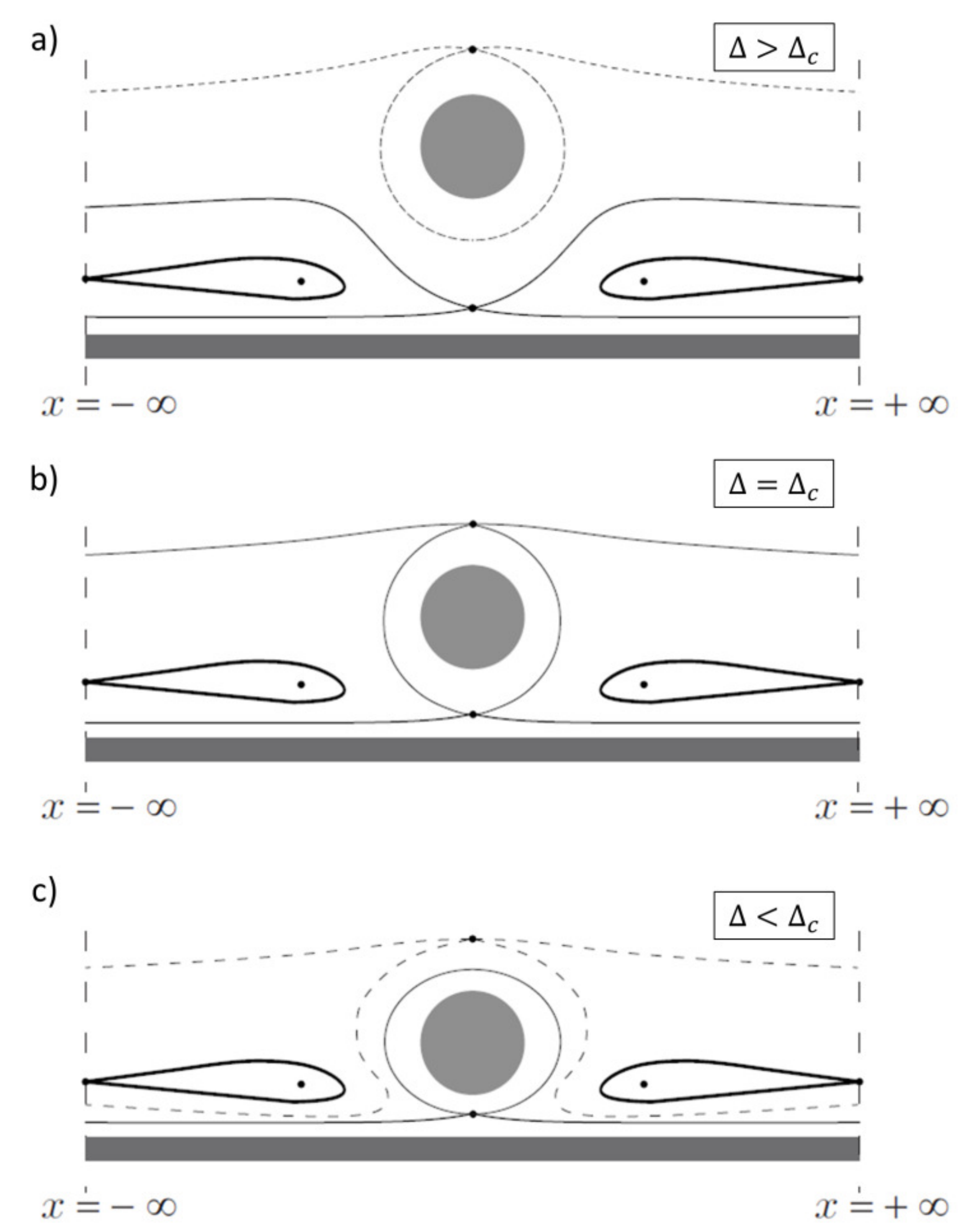}
\caption{Topological transition in phase space as the gap $\Delta$ is varied while the vortex circulation $\Gamma$ is kept fixed (and $\gamma=0$). Here only a schematic of the separatrices is shown.}
\label{fig:sepa3}
\end{figure}

A similar change in the topology occurs if one keeps the vortex circulation fixed and varies the gap $\Delta$ between the cylinder and the plane. This is illustrated in  Fig.~\ref{fig:sepa3} where we show a schematic of the separatrices for $\Gamma$ fixed, $\gamma=0$, and for three different values of $\Delta$.  In this figure we represent  the fixed points at $x=\pm \infty$ as visible points to indicate more clearly the   nilpotent saddle loops encircling the two centers. As seen in Fig.~\ref{fig:sepa3}(b), there is a critical  gap $\Delta_c$  for which a heteroclinic loop forms  connecting the two saddles on the normal axis, whereas for $\Delta>\Delta_c$ ($\Delta<\Delta_c$) this heteroclinic orbit becomes a homoclinic loop associated with the upper (lower) saddle.

\section{Discussion and Conclusions}
\label{sec:DC}

We have studied a point-vortex model for the formation of recirculating eddies in front of  a circular cylinder placed above a plane wall in the presence of a uniform stream. By mapping the fluid domain in the physical plane onto an annular region in an auxiliary complex plane, we have been able to compute an explicitly formula for the Hamiltonian of the system in terms of certain special functions related to elliptic theta functions. A detailed analysis of the equillibria of the system and their associated separatrices was presented. In particular, we have shown that there is a topological transition in phase space as  the gap between the cylinder and the plane or the vortex strength is varied, which implies different sets of  possible trajectories  for each topology.  

Special emphasis was given to the study of the equilibrium in front of the cylinder
and how it varies with the physical parameters of the model. 
It was shown that the vertical position of this equilibrium is more strongly dependent on the amount of circulation around the cylinder and less so on the gap  between the cylinder and the plane wall.
This property of the model  may  help to understand the behavior seen in the experiments  \cite{lin} where the center of the recirculating eddy moves away from  the plane boundary and towards the cylinder as the gap decreases.  Reducing the gap in the real case leads to a natural  increase in  circulation around the cylinder, and we have argued that  the latter  is the dominant effect on the position of the center of the recirculating eddy.

Our analysis  can in principle be extended to the case of vortex motion around multiple circular cylinders by means of the formalism of  the  Schottky-Klein prime functions. This formalism is well suited to construct conformal mappings between multiply connected domains of arbitrary connectivity and  was used  by Crowdy \cite{crowdy2005} to study the motion of a point vortex  around multiple circular cylinders with no background flow.  The inclusion of an imposed stream will likely give rise to several additional fixed points and produce a richer set of topological patterns in phase space.  This  will be pursued in more detail elsewhere.

\section{Acknowledgments}
This work was supported in part by the Brazilian agencies CNPq (grant 308290/2014-3) and CAPES, the latter through the PVE grant 88887.125151/2015-00 from the Science Without Borders program.  G.L.V.~also acknowledges support from the European Union under the Grant No.~PIRSES-GA-2013-612669. M.M. acknowledges support from the Research Council of Norway through its Centre of Excellence funding scheme with Project No. 262644. We are grateful to Prof.~Chang Lin for kindly providing us with the photographs shown in Fig.~\ref{fig:comparison}.

\end{document}